\documentclass[aps,prd,twocolumn,superscriptaddress,nofootinbib]{revtex4-1} 
\usepackage[utf8]{inputenc}
\usepackage{hyperref}
\hypersetup{
    colorlinks=true,
    linkcolor=blue,
    filecolor=magenta,      
    urlcolor=cyan,
    pdfpagemode=FullScreen,
}
\usepackage{verbatim}

\usepackage{graphicx}
\usepackage[dvipsnames]{xcolor}
\usepackage{soul,xcolor}
\usepackage{xcolor, ulem, color, framed}
\usepackage{amsmath}
\usepackage{amssymb}
\usepackage{textcomp}
\usepackage{physics}

\newcommand{\bt}{\mathbf{b}}
\newcommand{\Pt}{\mathbf{P}}
\newcommand{\kt}{\mathbf{k}}
\newcommand{\xt}{\mathbf{x}}
\newcommand{\rt}{\mathbf{r}}

\renewcommand{\d}{{\rm d}}
\newcommand{\snn}{\sqrt{s_{\rm NN}}}
\newcommand{\Pb}{\textrm{Pb}}

\begin{document}

\title{Constraining hot and cold nuclear matter properties from heavy-ion collisions and deep-inelastic scattering}

\author{Anton Andronic} 
\email{andronic@uni-muenster.de}
\affiliation{Institut f\"ur Kernphysik, Universit\"at M\"unster, D-48149 Münster, Germany}

\author{Nicolas Borghini} 
\email{borghini@physik.uni-bielefeld.de}
\affiliation{Fakult\"at f\"ur Physik, Universit\"at Bielefeld, D-33615 Bielefeld, Germany}

\author{Xiaojian Du} 
\email{xiaojian.du@usc.es}
\affiliation{Instituto Galego de Física de Altas Enerxías IGFAE, Universidade de Santiago de Compostela, E-15782 Galicia, Spain}

\author{Christian Klein-B\"osing} 
\email{christian.klein-boesing@uni-muenster.de}
\affiliation{Institut f\"ur Kernphysik, Universit\"at M\"unster, D-48149 Münster, Germany}

\author{Renata Krupczak} 
\email{rkrupczak@physik.uni-bielefeld.de}
\affiliation{Fakult\"at f\"ur Physik, Universit\"at Bielefeld, D-33615 Bielefeld, Germany}

\author{Hendrik Roch} 
\email{Hendrik.Roch@wayne.edu}
\affiliation{Department of Physics and Astronomy, Wayne State University, Detroit, Michigan 48201, USA}

\author{Sören Schlichting}
\email{sschlichting@physik.uni-bielefeld.de}
\affiliation{Fakult\"at f\"ur Physik, Universit\"at Bielefeld, D-33615 Bielefeld, Germany}

\begin{abstract}
We perform a global analysis of deep-inelastic $e+p$ scattering data from HERA and transverse energy distributions in $p+p$ and $p+\Pb$ collisions, alongside charged hadron multiplicities in $\Pb+\Pb$ collisions at $\sqrt{s_{\mathrm{NN}}} = 5.02\;\mathrm{TeV}$ from ALICE.
Using a saturation-based initial state model grounded in high-energy QCD, we determine the early-time non-equilibrium shear viscosity to entropy density ratio $\eta/s$ of the quark-gluon plasma. 
Our results provide new insights into the early-time transport properties of nuclear matter under extreme conditions. 
\end{abstract}

\date{\today}

\maketitle
\section{Introduction}
The study of the quark-gluon plasma (QGP) plays an important role in understanding the behavior of strongly interacting matter under extreme conditions.
High-energy heavy-ion collisions, such as those performed at facilities such as the Large Hadron Collider (LHC) and the Relativistic Heavy Ion Collider (RHIC), provide a unique opportunity to study this state of matter.
Key questions include the extraction of transport properties of the QGP, such as the shear viscosity to entropy density ratio $\eta/s$, which plays a crucial role in the dynamics governing the non-equilibrium and the hydrodynamics stages of the evolution of the fireball~\cite{Achenbach:2023pba,Arslandok:2023utm}.

Recent advances in theoretical models, particularly in the pre-equilibrium stage of evolution, allow for a more systematic connection between the initial-state physics and the final-state observables in heavy-ion collisions~\cite{schlichting2019first,Kurkela:2018wud,Giacalone:2019ldn}.
This study will provide the initial state of the collisions by a gluon saturation-based description within the Color Glass Condensate (CGC) effective field theory of high-energy QCD~\cite{Mantysaari:2015uca}.

In addition to theoretical modeling, measurements of deep inelastic $e+p$ scattering (DIS) at HERA~\cite{H1:2009pze} have provided critical constraints on the partonic structure of the proton at a small momentum fraction, $x$.
These constraints are particularly important for modeling the initial state of collisions involving protons and heavy nuclei, particularly at LHC energies, where gluon densities are large, and saturation effects dominate~\cite{Morreale:2021pnn}.

In this work, we perform a combined fit of data from deep-inelastic $e+p$ scattering, transverse energy distributions in $p+p$ and $p+\Pb$ collisions, and charged hadron multiplicities in $\Pb+\Pb$ collisions at $\snn=5.02\;\mathrm{TeV}$.
Using these complementary datasets combined with a saturation-physics-based initial state model, we aim to extract the early-time non-equilibrium $\eta/s$.
This approach not only bridges the gap between cold and hot nuclear matter but also provides a robust way to constrain QGP transport properties during its initial evolution.

In Section~\ref{sec:theoretical_background}, we introduce the saturation-physics-based model employed in this work and outline the procedure for computing initial conditions for the various collision systems analyzed. 
Section~\ref{sec:data_selection} discusses the selection of experimental data used to constrain the model parameters. 
In Section~\ref{sec:global_fit}, we detail the global fitting procedure, which incrementally incorporates additional model parameters. 
This section also covers the extraction of the effective pre-equilibrium temperature. 
Finally, in Section~\ref{sec:conclusions}, we summarize the key findings of the paper and compare our results for $\eta/s$ with those from other works.

\section{Theoretical description of $e+p$, $p+p$, $p+A$ and $A+A$ collisions}
\label{sec:theoretical_background}
We will describe hadronic reactions based on the Color Glass Condensate (CGC) effective field theory of high-energy QCD. Within this framework, elementary cross-sections and other inclusive observables in $e+p$, $p+p$, $p+A$, and $A+A$ can be formulated in terms of correlation functions of light-like Wilson lines $V_\xt$. 
Evidently, the simplest such correlation function is the (fundamental) dipole
\begin{align}
D_{\rm fun}(x,\rt,\xt)=\frac{1}{N_c}\expval{\text{tr}_{f} \left(V_{\xt+\rt/2} V^{\dagger}_{\xt-\rt/2}\right) }_{x},
\label{eq:DipoleFun}
\end{align}
which characterizes the scattering of a color dipole off a hadronic target~\cite{Mantysaari:2015uca}. 
In Eq.~\eqref{eq:DipoleFun} we adopted the standard notation $\expval{\dots}_{x}$ to denote the average at a given value of the longitudinal momentum fraction $x \ll 1$. 
We refer to Ref.~\cite{Garcia-Montero:2025hys} and references therein, for a detailed discussion of the connection with other hadronic structure functions.

\subsection{Structure of the nucleon and nucleus}
While different parametrizations of the dipole amplitude have been explored in the literature~\cite{Golec-Biernat:1998zce,Golec-Biernat:1999qor,Rezaeian:2012ji,Kowalski:2003hm}, within this work, we will stick to the simplest parametrization due to Golec-Biernat and W{\"u}sthoff (GBW):
\begin{align}
D_{\rm fun}(x,\rt,\xt)=\exp\left(-\frac{Q_{s,\rm fun}^2(x,\xt) \rt^2 }{4}\right),
\label{eq:DipoleFunGBW}
\end{align}
where $Q_{s,\rm fun}^2(x,\xt)$ is the (fundamental) saturation scale, which in addition to $x$ also depends on the transverse position $\xt$ inside the nucleon or nucleus. 
Although in the description of inclusive DIS data, it is often customary to neglect the transverse coordinate $(\xt)$ dependence, the latter is necessary to describe the energy deposition in hadronic reactions. 
We will, therefore, consistently employ the following standard parametrization 
\begin{align}
Q_{s,A}^{2}(x,\xt)= Q_{s,p}^{2}(x)~\sigma_{0} T_{A}(\xt)
\label{eq:QsAParam}
\end{align}
where $Q_{s,p}(x)$ is the $x$ dependent (average) saturation scale of the proton\footnote{Specifically one has $Q_{s,p}^{2}(x)=\sigma_{0}^{-1}\int \mathrm{d}^2\xt~Q_{s,p}^{2}(x,\xt)$.}, and $T_{A}(\xt)$  is the nuclear thickness obtained from a MC Glauber sampling of nucleon positions inside a nucleus~\cite{McLerran:2015qxa} according to
\begin{align}
T_{A}(\xt)&=\sum_{i\in A} T_p(\xt-\xt_i),\\ 
T_p(\xt)&=\frac{1}{2\pi B_{G}}e^{-\xt^2/(2B_{G})}\;,
\end{align}
where $B_{G}=4\; {\rm GeV}^{-2}$~\cite{Rezaeian:2012ji} is the nucleon size, and $\sigma_{0}=2\pi B_{G}$
such that the factor $\sigma_{0} T_{A}(\xt)$ in Eq.~(\ref{eq:QsAParam}) effectively counts the number of nucleons at a given transverse position $\xt$. 
Then the average saturation scale of the proton is parameterized as~\cite{Rezaeian:2012ji}
\begin{align}
Q_{s,p}^{2}(x) = Q_{s,0}^{2}~x^{-\lambda} (1-x)^{\delta},
\label{eq:p_saturation_scale}
\end{align}
where $Q_{s,0}^{2},\lambda$ and $\delta$ can be treated as parameters to be determined from the fit to the DIS data.

While the fundamental dipole distribution in Eq.~(\ref{eq:DipoleFun}) is sufficient to calculate the inclusive DIS cross section at leading order, we note that to describe energy deposition in hadronic collisions, we will also need the so-called unintegrated gluon distribution which is defined as
\begin{align}
\Phi_{(U)}(x,\kt,\xt)=\frac{\pi(N_c^2-1)}{g^2}D_{(U)}^{(1)} (x,\kt,\xt),
\end{align}
with $D_{(U)}^{(1)} (x,\kt,\xt)$ being the Fourier transform of the adjoint dipole distribution
\begin{align}
D_{(U)}^{(1)} (x,\kt,\xt) = \frac{\kt^2}{ N_c}\int{\rm d}^2\rt\;\frac{1}{N_c^2-1} \mathrm{tr} \left[ V_{\xt+\rt/2} ^{\rm adj} V_{\xt-\rt/2}^{\dagger,\rm adj} \right] e^{{\rm i}\kt \rt}
\end{align}
and $N_c$ denotes the number of colors.
By assuming Casimir scaling of the dipole distribution, the unintegrated gluon distribution in the GBW model is then given by
\begin{align}
\begin{split}
\Phi_{(U)}(x,\kt,\xt)=&\,4\pi^2\frac{N_c^2-1}{g^2 N_c}\frac{\kt^2}{ Q_{s, \rm adj}^2(x,\xt)}\\
&\times\exp \left( -\frac{\kt^2}{Q_{s,\rm adj}^2(x,\xt)} \right),
\end{split}
\end{align}
where $Q_{s,\rm adj}^2(x,\xt)=(C_A/C_F) Q_{s,\rm fun}^2(x,\xt)$ denotes the adjoint saturation scale.

Another key ingredient for the saturation model to describe charged hadron distributions in $p+p$ collisions is the implementation of fluctuations of the saturation scale $Q_{s,0}^2$ on the nucleon-by-nucleon level~\cite{Iancu:2004es,Marquet:2006xm,McLerran:2015qxa}.
These fluctuations are assumed to follow log-normal statistics~\cite{McLerran:2015qxa}:
\begin{align}
    p(\mathcal{Z}) = \frac{1}{\sqrt{2\pi}\sigma_Q}\exp\left(-\frac{\mathcal{Z}^2}{2\sigma_Q^2}\right)
\label{eq:log_norm_Qs_fluct}
\end{align}
with $\mathcal{Z}=\ln(Q_{s,0}^2/\langle Q_{s,0}^2\rangle)$ and $\sigma_Q=0.5$.

\subsection{Inclusive DIS cross-section}
When considering deep-inelastic $e+p$ scattering, we will follow Ref.~\cite{Rezaeian:2012ji} and consider the inclusive $\gamma^{*}p$ reduced cross-section $\sigma_r$ as a function of the kinematic variables $x$, $y$ and $Q^2$. The starting point for the calculation is the leading order cross-sections for the scattering of a longitudinally $(L)$ or transversely $(T)$ polarized virtual photon given by
\begin{align}
\sigma^{\gamma^*p}_{T,L}&(x,Q^2)= \\
& 2\int {\rm d}^2\xt\, {\rm d}^2\rt\int_0^1 {\rm d}z |\Psi_{T,L}(r,z,Q^2)|^2\mathcal{N}(x,\rt,\xt) \nonumber
\end{align}
where $\mathcal{N}(x,\rt,\xt)=1-D_{\rm fun}(x,\rt,\xt)$ is the dipole scattering amplitude and the longitudinal and transversely polarized photon wave functions take the forms
\begin{align}
&|\Psi_{T}(r,z,Q^2)|^2=\frac{N_c\alpha_{\rm em}}{2\pi^2}\times 
\nonumber\\ 
& \quad \sum_f e_f^2\Big\{[z^2+(1-z)^2]\bar{Q}^2K_1(\bar{Q}r)^2+m_f^2K_0(\bar{Q}r)^2\Big\},
\end{align}
and
\begin{align}
&|\Psi_{L}(r,z,Q^2)|^2=
\nonumber\\
& \quad\frac{N_c\alpha_{\rm em}}{2\pi^2}\sum_f e_f^2\left\{4Q^2z^2(1-z)^2K_0(\bar{Q}r)^2\right\},
\end{align}
with $\bar{Q}^2=z(1-z)Q^2+m_f^2$, and the modified Bessel functions $K_0$ and $K_1$.
Here $z$ denotes the fraction of the light cone momentum of the virtual photon carried by the quark, $m_f$ is the quark mass, $e_f$ is the electric charge of the quark with flavor $f$, and $\alpha_{\rm em}$ denotes the electromagnetic fine structure constant.\footnote{We include the first three light quark flavors with $m_f=0.14\;\mathrm{GeV}$~\cite{Golec-Biernat:1998zce}.}
The proton structure function $F_2$ and longitudinal structure function $F_L$ are then given by
\begin{align}
&F_2(x,Q^2)=\frac{Q^2}{4\pi^2\alpha_{\rm em}}\left[\sigma^{\gamma^*p}_{L}(x,Q^2)+\sigma^{\gamma^*p}_{T}(x,Q^2)\right],\\
&F_L(x,Q^2)=\frac{Q^2}{4\pi^2\alpha_{\rm em}}\left[\sigma^{\gamma^*p}_{L}(x,Q^2)\right],
\end{align}
and the reduced cross-section $\sigma_r$ which is a directly measurable observable is expressed in terms of the inclusive $F_2$ and $F_L$ by
\begin{align}
\sigma_r(x,y,Q^2)= F_2(x,Q^2)-\frac{y^2F_L(x,Q^2)}{1+(1-y)^2}
\end{align}
resulting in
\begin{align}
\begin{split}
\sigma_r&(x,y,Q^2) = \frac{Q^2}{4\pi^2\alpha_{\rm em}}\times\\
&\left[\left(\frac{2-2y}{1+(1-y)^2}\right)\sigma^{\gamma^*p}_{L}(x,Q^2)+\sigma^{\gamma^*p}_{T}(x,Q^2)\right]\;.
\end{split}
\end{align}

\subsection{Initial state in hadronic collisions}
\label{subsec:HIC_initial_condition}
Within the CGC effective theory of high-energy QCD, the initial energy deposition in heavy-ion collisions can be computed from the solutions of the classical Yang-Mills equations~\cite{Krasnitz:1999wc,Lappi:2003bi,Schenke:2012wb}.
When saturation effects in one of the colliding nuclei can be ignored 
the spectrum of produced gluons exhibits $k_{\rm T}$ factorization~\cite{Lappi:2017skr,Blaizot:2010kh}, and can be calculated according to\footnote{See e.g. Eq.~(4.40) in Ref.~\cite{Lappi:2017skr} and Eq.~(29) in~\cite{Blaizot:2010kh}.} the spectrum of initially produced gluons per unit transverse area, transverse momentum $\Pt$ of the gluons, and momentum rapidity $Y$, ${\rm d}N_{g}/({\rm d}^2\xt\, {\rm d}^2\Pt\, {\rm d}Y\, {\rm d}y)$, in hadronic collisions
\begin{align}
\label{eq:NgSpectra}
\begin{split}
&\frac{{\rm d}N_{g}}{{\rm d}^2\xt\, {\rm d}^2\Pt\, {\rm d}Y\, {\rm d}y} 
=\frac{\alpha_s N_c}{\pi^4 \Pt^2 (N_c^2-1)} \delta(Y-y) \\
&\quad \int\!\frac{{\rm d}^2\kt}{{ (2\pi)^2}}\Phi_{A}\!\left(x_A,\xt+\frac{\bt}{2},\kt\right) \Phi_{B}\!\left(x_B,\xt-\frac{\bt}{2},\Pt -\kt\right),
\end{split}
\end{align}
where in leading order kinematics $x_{A/B}=|\Pt|\,{\rm e}^{\pm Y\!}/\sqrt{s_{\rm NN}}$, $\Phi_{A/B}$ denotes the unintegrated gluon distributions (UGDs) of the colliding nuclei $A$ and $B$, and $\bt$ is the impact parameter of the collision.
Numerical investigations in Refs.~\cite{Blaizot:2010kh,Schlichting:2019bvy} show that higher-order saturation corrections to Eq.~(\ref{eq:NgSpectra}) are typically small, in particular when considering high momenta.

By integrating the gluon spectrum over transverse momenta $\Pt$, the initial transverse energy density per unit rapidity is then obtained as 
\begin{align}
[e(\xt)\tau]_{0} 
=K\int\!{\rm d}Y\int\!{\rm d}^2\Pt\, |\Pt|\frac{{\rm d}N_{g}}{{\rm d}^2\xt\, {\rm d}^2\Pt\, {\rm d}Y\, {\rm d}y}.
\label{eq:energyperrapidity}
\end{align}
As Eq.~(\ref{eq:NgSpectra}) is valid to leading order, we allow for a multiplicative $K$-factor of order unity to account for higher-order corrections. This global scale $K$ in the energy is an additional free parameter in the model, which we shall determine hereafter.

Evaluating the integrals for the GBW model, by approximating the $x$ dependence of the UGDs in a similar way to the IP-Glasma model~\cite{Schenke:2012hg,Schenke:2013dpa} as 
\begin{align}
\label{eq:QsXSelfConsistent}
x_{A/B} = \frac{Q_{s,A/B}(x_{A/B},\xt)\,{\rm e}^{\pm Y}}{\snn}
\end{align}
one then obtains the initial energy per unit rapidity as~\cite{Borghini:2022iym,Garcia-Montero:2023gex}
\begin{align}
\begin{split}
\label{eq:e0}
[e(\xt)\tau]_{0} =&\,K\frac{N_c^2-1}{4g^2 N_c\sqrt{\pi}} \frac{Q_{s,A}^2 Q_{s,B}^2}{(Q_{s,A}^2+Q_{s,B}^2)^{5/2}} \\
&\times\left(2Q_{s,A}^4+7Q_{s,A}^2Q_{s,B}^2+2Q_{s,B}^4\right)\;,    
\end{split}
\end{align}
with $Q_{s,A/B}(x,\xt)$ self-consistently determined from Eqs.~\eqref{eq:QsAParam} and \eqref{eq:QsXSelfConsistent}.
In Appendix~\ref{app:self_consistent_Qs} we introduce an additional proportionality constant in Eq.~\eqref{eq:QsXSelfConsistent} to take into account that the $x$ dependence of the UGDs is only approximated.

When considering minimum bias $p+p$ and $p+A$ collisions, we will assume that due to the absence of strong final state effects, the initial transverse energy per unit rapidity
\begin{align}
\frac{{\rm d}E_{\bot}}{{\rm d}y}=\int\!{\rm d}^2\xt\, [e(\xt)\tau]_{0} 
\end{align}
can be directly compared to the final state energy per unit rapidity ${\rm d}E_{\bot}/{\rm d}y$ measured in experiments.

\section{Experimental Data Selection}
\label{sec:data_selection}
We summarize the model parameters and the steps in our fitting procedure where they appear in Tab.~\ref{tab:model_parameters_list}.
\begin{table}[!tb]
    \caption{\label{tab:model_parameters_list}Model parameters and the collision system where they are relevant. Parameters that are fixed in our analysis are shown with (\checkmark).}
    \centering
    \begin{tabular}{c|c|c|c}
    \hline\hline
    Parameter & $\gamma^*+p$ & $p+p$/$p+A$ & $A+A$ \\
    \hline
    $Q_{s,0}$ & \checkmark & \checkmark & \checkmark \\
    $\lambda$ & \checkmark & \checkmark & \checkmark \\
    $\delta$ & (\checkmark) & (\checkmark) & (\checkmark) \\
    $\sigma_Q$ & (\checkmark) & (\checkmark) & (\checkmark) \\
    $K$ & & \checkmark & \checkmark \\
    $\eta/s$ & & & \checkmark \\
    \hline\hline
    \end{tabular}
\end{table}
The table nicely shows that by going from smaller to larger collision systems, we can fix the parameters in a multi-step process and finally determine the ratio $\eta/s$ with the heavy-ion collision dataset.

For the DIS fit, we select inclusive $\gamma^*+p$ scattering cross-section data from HERA~\cite{H1:2009pze} within the kinematic range relevant for high-energy heavy-ion collisions, namely $2\;\mathrm{GeV^2}\leq Q^2\leq 22\;\mathrm{GeV^2}$ and $x\leq 0.01$.
This selection results in 100 experimental data points used in this part of the fit.

To determine the $K$ factor for the initial state of the heavy-ion collision, we utilize measurements of charged hadron multiplicity and mean transverse momentum from minimum bias $p+p$ and $p+\Pb$ collisions, as reported by the ALICE collaboration at $\snn=5.02\;\mathrm{TeV}$~\cite{ALICE:2022xip}.
The kinematic range of these datasets is $\abs{\eta}<0.8$ and $0.15\;\mathrm{GeV}<p_{\rm T}< 10\;\mathrm{GeV}$.
For both systems, we use the experimental measurements of the multiplicity and mean transverse momentum extrapolated to zero transverse momentum to compute $\langle\d E_\perp/\d y\rangle$.
Further details on the transformation and the zero-$p_{\rm T}$ extrapolation are provided in Appendix~\ref{app:ET_dist}.

Finally, the fit of the shear viscosity to entropy density ratio $\eta/s$, is performed using the charged hadron yield at midrapidity, $\d N_{\rm ch}/\d \eta$, in the most central centrality class, as measured by the ALICE Collaboration in $\Pb-\Pb$ collisions at $\snn=5.02\;\mathrm{TeV}$~\cite{ALICE:2015juo}.

\section{Global fit of DIS, $p+p$/$p+A$ and $A+A$ data}
\label{sec:global_fit}
In this section, we outline the steps required to determine the model parameters needed to extract the pre-equilibrium value of $\eta/s$ and the effective temperature during this phase.

\subsection{DIS fit}
\label{subsec:DIS_fit}
The DIS cross-section values in the model calculations are influenced by four model parameters, see Table~\ref{tab:model_parameters_list}.
The two most important ones for our fit are parameters of the average saturation scale~\eqref{eq:p_saturation_scale}: the value of $Q_{s,0}$ and the parameter $\lambda$ governing the growth of the saturation scale with decreasing $x$.
We fix the parameter $\delta\equiv 1$, which models the decrease of the saturation scale with increasing $x$, since our fit is restricted to the kinematic region relevant for heavy-ion collisions, $x\leq 0.01$.
The parameter $\sigma_Q$ from Eq.~\eqref{eq:log_norm_Qs_fluct}, which controls the fluctuation of the saturation scale, is present in the description of both DIS cross-sections and $p+p/p+A$ systems to describe the charged hadron spectrum.
In this analysis, we keep this parameter fixed at $\sigma_Q\equiv 0.5$, following Ref.~\cite{McLerran:2015qxa}.

By fixing $\delta$ and $\sigma_Q$, we are left with two free parameters -- $Q_{s,0}$ and $\lambda$ -- which we determine from the HERA data.
To extract these parameters, we perform a Bayesian analysis, generating 1000 training points in a Maximum Projection Latin Hypercube Design~\cite{doi:10.1080/00224065.2019.1611351,10.1093/biomet/asv002} to efficiently cover the parameter space.
We assume uniform priors in the ranges $[0.1,0.8]\;\mathrm{GeV}$ for $Q_{s,0}$ and $[0.1,0.5]$ for $\lambda$.
The full model is then used to compute the cross-sections at all training points.
We vary the model uncertainty such that the reduced $\chi^2$ is approximately unity, which is achieved by assuming a model uncertainty of $6.6\%$.
These results are used to train a Gaussian Process (GP) emulator.
For this, we employ the PCSK emulator from the Python package \texttt{surmise}~\cite{surmise2023}, developed by the BAND collaboration, to replace the full model with an efficient surrogate.
To validate the emulator predictions, we use metrics introduced in Ref.~\cite{Roch:2024xhh}, which show that the root-mean-square uncertainty of the emulator is below $0.6\%$.

With the trained emulator, we perform a Markov Chain Monte Carlo (MCMC) analysis to determine the posterior distributions of the parameters.
To be consistent with the experimental data, we only consider the diagonal elements of the emulator's covariance matrix in the MCMC step, and the model uncertainty is added to the emulator uncertainty to reflect the systematic model uncertainty.
This diagonal approximation neglects potential correlations among nearby DIS data points.
Their impact is partially absorbed into the effective model uncertainty, and the resulting parameter constraints should therefore be interpreted as conditional on this approximation.

We note that a $6.6\%$ model uncertainty applied at the level of the DIS cross sections does not directly translate into percent-level uncertainties on the fitted parameters.
Since the observables depend in a correlated and sensitive manner on $Q_{s,0}$ and $\lambda$ across a large number of data points, the combined constraints from the full dataset can still determine these parameters at the $\mathcal{O}(1\%)$ level.
Specifically, we use the MCMC sampler package \texttt{pocoMC}~\cite{Karamanis:2022alw,Karamanis:2022ksp}.\footnote{For further details on the GP emulator and MCMC, see Refs.~\cite{Roch:2024xhh,Jahan:2024wpj,Jahan:2025cbp}. The emulator and MCMC tools are available at Ref.~\cite{hendrik_roch_2024_12807892}.}

\begin{figure}[!tb]
    \centering
    \includegraphics[width=\linewidth]{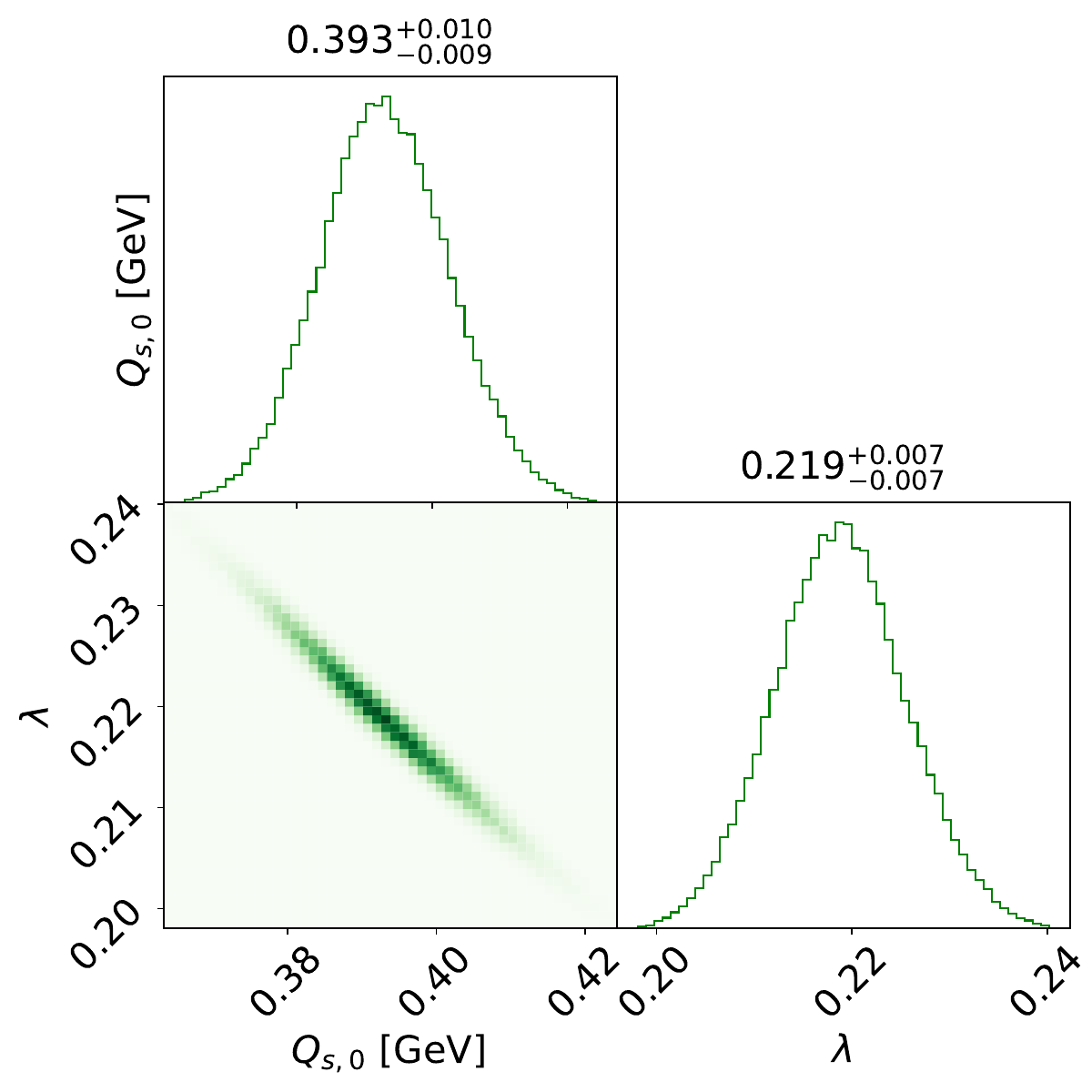}
    \caption{Posterior distributions for the parameters $Q_{s,0}$ and $\lambda$ in the DIS fit. The values above the plots indicate the median and the $1\sigma$ interval.}
    \label{fig:DIS_corner}
\end{figure}

The resulting posterior distributions are shown in Fig.~\ref{fig:DIS_corner}, revealing a clear anti-correlation between the two parameters.
We extract the maximum a posteriori probability (MAP) parameter set from these distributions, yielding $Q_{s,0}=0.393\;\mathrm{GeV}$ and $\lambda=0.219$. 

\begin{figure*}[!tb]
    \centering
    \includegraphics[width=\linewidth]{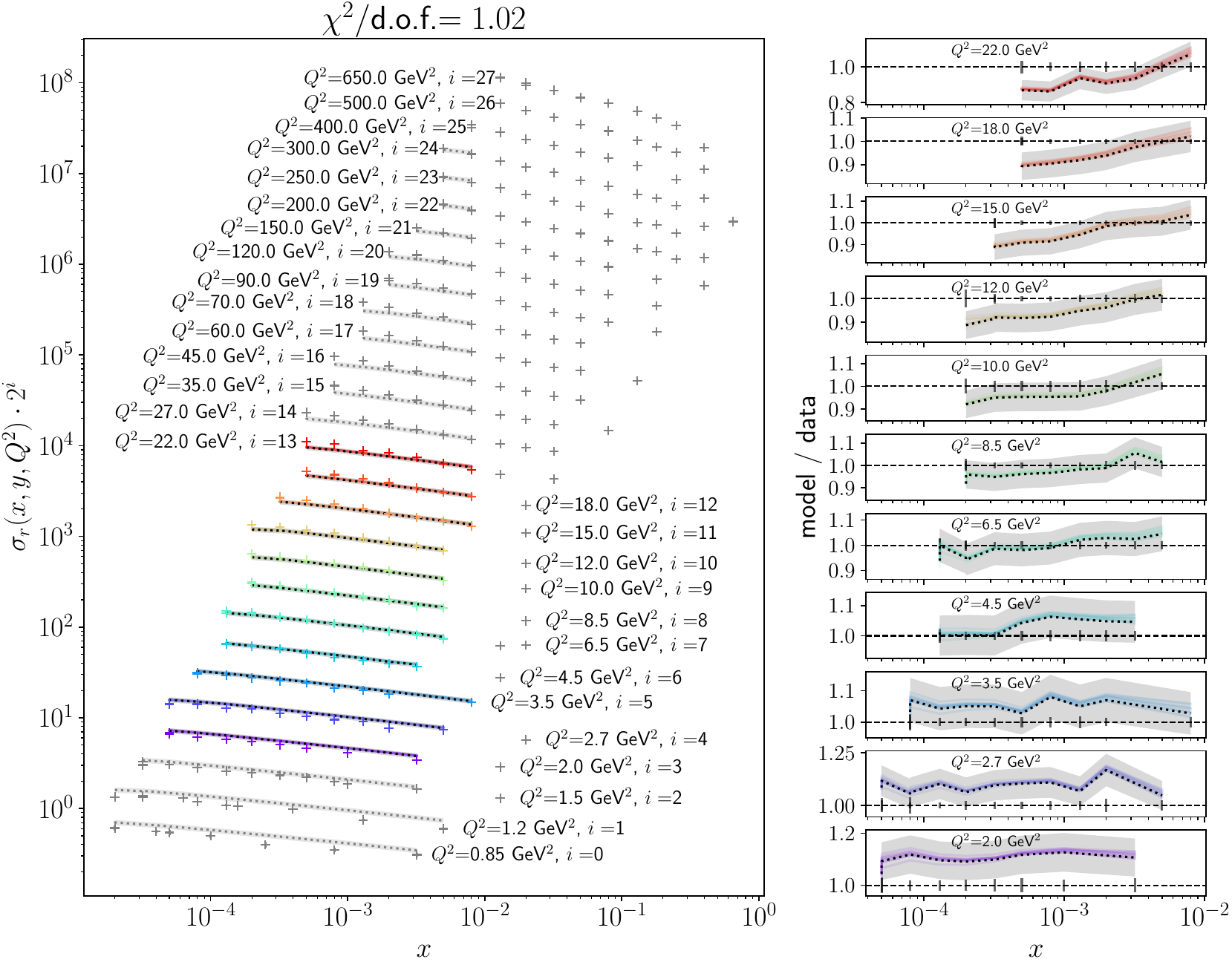}
    \caption{(Left) Cross-section predictions using the MAP parameters (black dotted line) and ten posterior sample predictions (colored lines), compared to the HERA data~\cite{H1:2009pze} in the relevant kinematic region as a function of $x$. All gray points are not included in the fit, and the dotted lines are model simulations at the MAP parameters.
    (Right) Model-to-data ratios for the MAP parameters (black dotted line) with 1$\sigma$ model uncertainty band for the $Q^2$ values used for the fit. The colored lines show the mean values of the posterior sample predictions (without uncertainty bands).}
    \label{fig:DIS_posterior_samples}
\end{figure*}

After determining the optimal parameters for the given model and dataset, we compare the results with experimental data.
Figure~\ref{fig:DIS_posterior_samples} shows the reduced cross sections $\sigma_r$ for all data points used in the fit with colored markers.
The colored lines represent ten posterior samples generated from the full model, and the black dotted line represents a full model run at the MAP parameters.
The full model calculation, when compared to the experimental data, yields a reduced $\chi^2$ of 1.02, indicating a statistically consistent description of the data given the combined experimental, emulator, and model uncertainties. This consistency is clearly reflected in the model-to-data ratios shown in the right panel of the figure.
We also find that the posterior sample predictions align well with the MAP predictions.
Data points corresponding to $Q^2$ values excluded from the fit, as well as those with $x>0.01$, are shown in gray.
The gray dotted lines represent MAP parameter predictions from the full model run, which reasonably describe the data at the excluded virtualities.

\subsection{Combined $p+p$/$p+A$ fit}
\label{subsec:pp_pA_fit}

To determine the energy scaling $K$-factor that accounts for higher-order corrections in the initial-state model, we use the probability distribution of the transverse energy per unit rapidity in minimum bias $p+p$ and $p+\Pb$ collisions obtained through the procedure described in Appendix~\ref{app:ET_dist}.
Calculating the mean transverse energy from these distributions yields $\langle\d E_\perp/\d y\rangle=6.80\pm 0.08\;\mathrm{GeV}$ for $p+p$ collisions and $23.2\pm 0.3\;\mathrm{GeV}$ for $p+\Pb$ collisions at $\snn=5.02\;$TeV.
Based on these results, we compute separate $K$-factors for the two systems and combine them to obtain the average value
\begin{equation}
\label{eq:K-value}
    K=1.93\pm 0.04,   
\end{equation}
which is used consistently in all subsequent analysis steps, including the $\Pb+\Pb$ calculations.
This is justified because the higher-order corrections captured by the $K$-factor are expected to be independent of the collision system.

\begin{figure}[!tb]
    \centering
    \includegraphics[width=\linewidth]{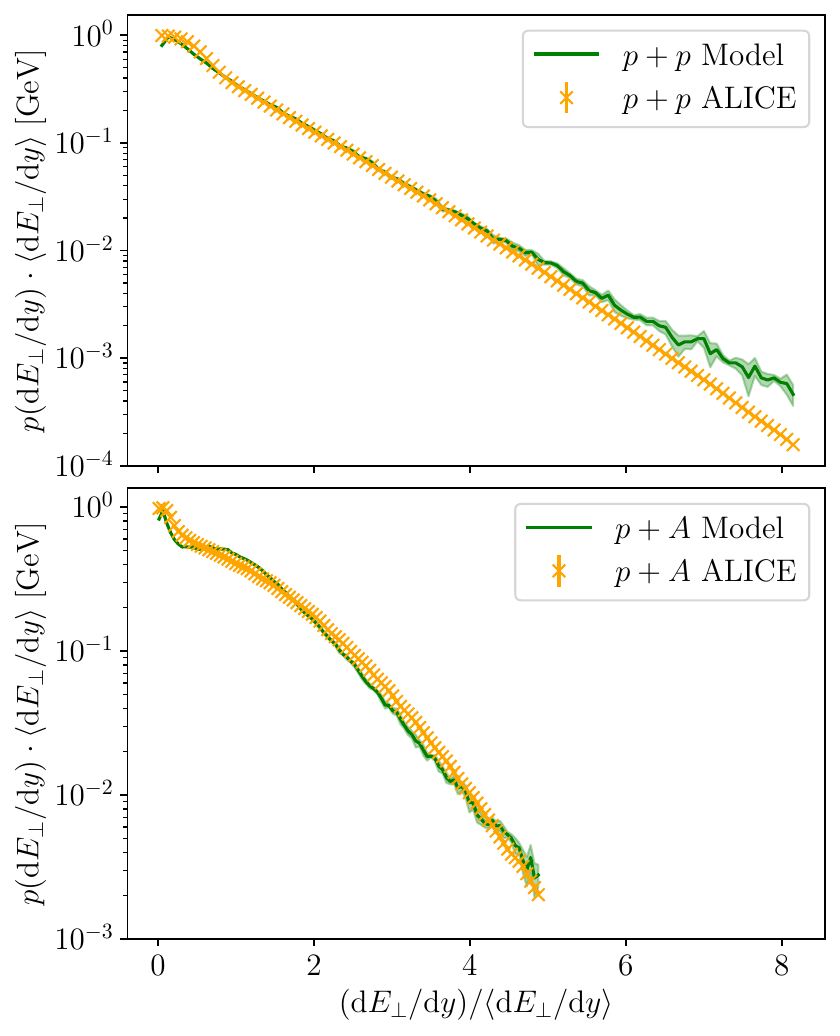}
    \caption{Probability distribution of the transverse energy per unit rapidity rescaled by the mean value. Orange points represent the distributions obtained using the procedure detailed in Appendix~\ref{app:ET_dist}, while the green line (with uncertainties) shows the saturation model results using the MAP parameters and ten posterior samples from the DIS fit.}
    \label{fig:dEdy_dist}
\end{figure}

Figure~\ref{fig:dEdy_dist} compares the scaled probability distributions for the transverse-energy rapidity density derived from the ALICE data~\cite{ALICE:2022xip} with the results of $2^{18}$ initial configurations from the saturation model.
As mentioned above, the $K$-factor has been fitted only to match the average values of the two distributions, not their full shapes.
To take into account the variations in the DIS parameters $Q_{s,0}$ and $\lambda$, we average the $K$-factors for the two systems over the MAP parameters and ten parameter sets sampled from the posterior distributions in Fig.~\ref{fig:DIS_corner}.  
For the $p+p$ system, the data and the model show good agreement up to approximately twice the mean value, indicating that the model captures the transverse energy distribution well for most events.
In the $p+\Pb$ system, the data and model match rather well below the mean transverse energy, but the model flattens around the mean value and exhibits a steeper falloff at higher transverse energies compared to the ALICE data.
We note that the shape of the distribution in $p+\Pb$ collisions can likely be improved by introducing additional impact parameter dependence, e.g.\ due to sub-nucleonic fluctuations (hot spots)~\cite{Eremin:2003qn,Loizides:2016djv,Garcia-Montero:2025bpn}. 
However, for the sake of simplicity these effects are not included in the present model. 
Nevertheless, the overall agreement between the model distributions and experimental data is sufficient to proceed with the fitting procedure under this setup.

\subsection{$A+A$ fit}
\label{subsec:AA_fit}

When turning to heavy-ion collisions, final-state effects become significant, making it advantageous to use charged particle multiplicity as a measure of the total entropy of the thermalized system.
During the space-time evolution of a heavy-ion collision, the initial energy density of the non-equilibrium system is converted to thermal entropy during the pre-equilibrium phase.
Based on Ref.~\cite{Giacalone:2019ldn}, the resulting entropy density of the near-equilibrium QGP can be expressed in terms of the initial transverse energy density as
\begin{align}
\label{eq:dS/dV}
\frac{{\rm d}S}{{\rm d}^2\xt\,{\rm d}y}=\frac{4}{3}C_{\infty}^{\tfrac{3}{4}}\left(\frac{\pi^2}{30}\nu_{\rm eff}\right)^{\!\tfrac{1}{3}}\left(4\pi\frac{\eta}{s}\right)^{\!\tfrac{1}{3}}\left[e(\xt)\tau\right]_0^{\tfrac{2}{3}},
\end{align}
where $C_\infty=0.87$ is a non-equilibrium constant, and $\nu_{\rm eff}$ represents the effective number of degrees of freedom in the QGP.
During the hydrodynamic expansion, the total entropy per unit rapidity, $\d S/\d y = \int\d^2\xt\,\d S/(\d y\,\d^2\xt)$, is approximately conserved and converted into charged particles at freeze out.
This makes the final-state charged hadron multiplicity particularly sensitive to the early-time viscosity of the QGP.
Equation~\eqref{eq:dS/dV} allows one to estimate the charged particle multiplicity $\d N_{\rm ch}/\d\eta$ as
\begin{align}
    \frac{\d N_{\rm ch}}{\d\eta} = \frac{4}{3}\frac{N_{\rm ch}}{S}C_\infty^{\tfrac{3}{4}}\left(4\pi\frac{\eta}{s}\right)^{\!\tfrac{1}{3}}\left(\frac{\pi^2}{30}\nu_{\rm eff}\right)^{\!\tfrac{1}{3}} \!\int\!\d^2\xt\; [e(\xt)\tau]_0^{\tfrac{2}{3}},
    \label{eq:mult_estimator}
\end{align}
where $S/N_{\rm ch}\equiv (\d S/\d y)/(\d N_{\rm ch}/\d y)$ is fixed to 7.5, based on the entropy per charged particle at freeze-out~\cite{Hanus:2019fnc}.
The effective number of degrees of freedom in the QGP is kept constant at $\nu_{\rm eff}=40$.
Equation~\eqref{eq:mult_estimator} directly determines the final-state multiplicity from the initial-state energy density without running a whole dynamical evolution, thus enabling efficient event processing for centrality selection.

To account for physical effects not captured in Eq.~\eqref{eq:mult_estimator}, such as higher-order $\eta/s$ corrections, we modify this equation by introducing a viscosity-dependent proportionality factor:
\begin{align}
    \frac{\d N_{\rm ch}}{\d\eta}\Big\vert_{\rm sim}=C\left(\frac{\eta}{s}\right)\cdot\frac{\d N_{\rm ch}}{\d\eta}\Big\vert_{\rm est},
    \label{eq:est_calib}
\end{align}
where the subscript ``sim'' represents multiplicities from state-of-the-art dynamical simulations of heavy-ion collisions, and ``est'' represents the estimation from Eq.~\eqref{eq:mult_estimator}.
To estimate $C(\eta/s)$, the dynamical evolution is performed for a small sample of initial states to ``calibrate" Eqs.~\eqref{eq:mult_estimator}--\eqref{eq:est_calib} to the output of full simulations, thereby enabling the subsequent use of the estimator without the need for additional event-by-event simulations.
For this calibration, full event simulations are conducted using a hybrid code setup~\cite{hendrik_roch_2024_12694840} that evolves saturation model initial conditions through the K{\o}MP{\o}ST~\cite{Kurkela:2018vqr} pre-equilibrium stage, followed by dissipative fluid-dynamical evolution in MUSIC~\cite{Schenke:2010rr,Schenke:2010nt,Paquet:2015lta} with particlization via iSS~\cite{Shen:2014vra} and subsequent evolution in the hadronic afterburner SMASH~\cite{SMASH:2016zqf}.
Since we are primarily interested in the total entropy of the system, the pre-equilibrium state in K{\o}MP{\o}ST is matched to the hydrodynamic phase using the equilibrium entropy-matching method from Ref.~\cite{Borghini:2024kll}.

\begin{figure}[!tb]
    \centering
    \includegraphics[width=\linewidth]{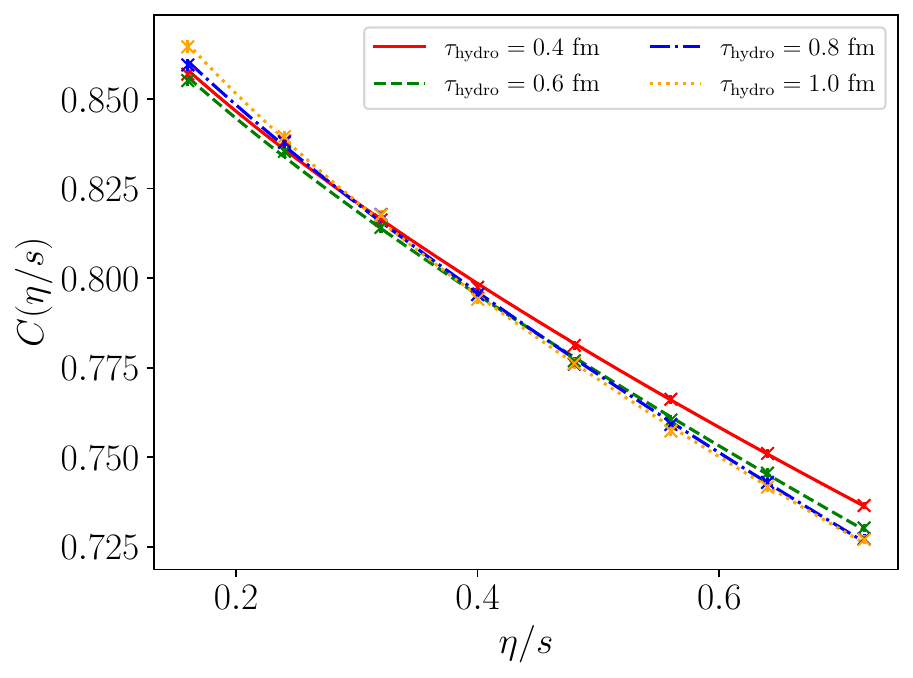}
    \caption{Proportionality factor $C(\eta/s)$ from Eq.~\eqref{eq:est_calib} (points) and the fitted parametrization (lines) from Eq.~\eqref{eq:C_estimator_prefac}, shown for four different $\tau_\mathrm{hydro}$ values.}
    \label{fig:C_calib_estimator}
\end{figure}

The calibration of the multiplicity estimator~\eqref{eq:mult_estimator}--\eqref{eq:est_calib} involves the previously determined average $K$-factor of the initial state [Eq.~\eqref{eq:K-value}] and four Gaussian samples around the mean, with the MAP parameters determined from the DIS fit.
Ten event-by-event dynamical simulations are performed at $b=0$ impact parameter with ten oversamplings in the Cooper--Frye procedure to compute the mean multiplicity and standard deviation.
These are carried out for several switching times $\tau_\mathrm{hydro}=\lbrace 0.4,0.6,0.8,1.0\rbrace\;\mathrm{fm}$ between K{\o}MP{\o}ST and MUSIC, and for $\eta/s$ values ranging from 0.16 to 0.72 in steps of 0.08.\footnote{This shows one strength of our approach since only 320 event-by-event simulations for each of the five parameter sets ($\langle K\rangle$ and four Gaussian samples) are performed to determine $C(\eta/s)$ for the different values of $\tau_{\rm hydro}$.}
These $\eta/s$ values are consistently used in the pre-equilibrium and hydrodynamic stages, while bulk viscosities are not considered in this study.
Charged hadron multiplicities at midrapidity are analyzed from the SMASH output using the Python package \texttt{SPARKX}~\cite{hendrik_roch_2024_14516063,Sass:2025opk}.
A linear fit determines the proportionality factor, displayed as points in Fig.~\ref{fig:C_calib_estimator}, for each $\tau_\mathrm{hydro}$ and $\eta/s$ in Eq.~\eqref{eq:est_calib}.
The calibration factor $C(\eta/s)$ is parametrized as
\begin{align}
C\!\left(\frac{\eta}{s}\right) = a +b\sqrt{\frac{\eta}{s}} + c \frac{\eta}{s},
    \label{eq:C_estimator_prefac}
\end{align}
which is a simple nonlinear ansatz that nicely fits the eight computed points from the full K{\o}MP{\o}ST+MUSIC+SMASH simulation setup with only three parameters, as shown in Fig.~\ref{fig:C_calib_estimator}.
The fit parameters for different $\tau_{\rm hydro}$ values are listed in Tab.~\ref{tab:C_parameters_est_calib}.
Note that the parameters $a,b,c$ in Eq.~\eqref{eq:C_estimator_prefac} are empirical fits for the calibration of the estimator~\eqref{eq:mult_estimator}--\eqref{eq:est_calib} and do not count as free model parameters.
This also holds for $C_\infty$, $\nu_{\rm eff}$ and $N_{\rm ch}/S$ which appear as separate constants in Eq.~\eqref{eq:mult_estimator}, but are altogether ``ingredients'' of the multiplicity estimator. 
\begin{table}[!tb]
    \caption{\label{tab:C_parameters_est_calib}Fit parameters for Eq.~\eqref{eq:C_estimator_prefac} at different $\tau_{\rm hydro}$ values.}
    \centering
    \begin{tabular}{c|c|c|c}
        \hline\hline
        $\tau_{\rm hydro}$ & $a$ & $b$ & $c$ \\
        \hline
        $0.4\;\mathrm{fm}$ & $0.94\pm 0.01$ & $-0.18\pm 0.04$ & $-0.07\pm 0.03$ \\
        $0.6\;\mathrm{fm}$ & $0.94\pm 0.01$ & $-0.16\pm 0.03$ & $-0.09\pm 0.02$ \\
        $0.8\;\mathrm{fm}$ & $0.95\pm 0.02$ & $-0.17\pm 0.04$ & $-0.10\pm 0.03$ \\
        $1.0\;\mathrm{fm}$ & $0.98\pm 0.01$ & $-0.26\pm 0.04$ & $-0.04\pm 0.03$ \\
        \hline\hline
    \end{tabular}
\end{table}

By combining Eqs.~\eqref{eq:mult_estimator}--\eqref{eq:C_estimator_prefac} the event-by-event multiplicity is then determined as
\begin{align}
    \frac{\d N_{\rm ch}}{\d\eta}\Big\vert_{\rm sim}&=C\left(\frac{\eta}{s}\right)\cdot\frac{4}{3}\frac{N_{\rm ch}}{S}C_\infty^{\tfrac{3}{4}}\left(4\pi\frac{\eta}{s}\right)^{\tfrac{1}{3}}\left(\frac{\pi^2}{30}\nu_{\rm eff}\right)^{\tfrac{1}{3}}\nonumber\\
    &\times\int\!\d^2\xt\; [e(\xt)\tau]_0^{\tfrac{2}{3}}
    \label{eq:est_calib2}
\end{align}
which allows for a highly efficient calculation. 
Before proceeding to the statistical analysis, we note that the sensitivity to the viscosity enters directly through the power dependence of $\eta/s$  as well as indirectly through the mild variation of $C(\eta/s)$. 
Because $\d N_{\rm ch}/\d\eta$ depends both on $K$, which is included in $[e(\xt)\tau]_0$, and $\eta/s$, it is crucial to constrain $K$ independently using $p+p$ and $p+\Pb$ data before fitting $\eta/s$ in $\Pb+\Pb$.

When performing the statistical analysis for $\Pb+\Pb$ collisions, we generate $2^{16}$ initial-state configurations for each parameter set to determine centrality classes using the \texttt{SPARKX} package and compare the most central class (0-2.5\%) to experimental measurements from the ALICE Collaboration.\footnote{Only the most central collisions are considered because the estimator calibration was performed at $b=0$ impact parameters. Additionally, the K{\o}MP{\o}ST+MUSIC setup is well-established in this regime, unlike in peripheral collisions, where the applicability is less certain.}
During the fitting process, we vary $\eta/s$ in the range $[0.16, 0.72]$ in 200 steps, compute the $\chi^2$, and identify the minimum.
To account for uncertainties from earlier steps, this procedure is repeated for combinations of all five values (average value and four Gaussian samples) of the $K$-factor, all eleven sets (MAP value and ten posterior samples) of DIS parameters, and four Gaussian samples of the parameters in $C(\eta/s)$.
This results in a total of 220 $\eta/s$ values obtained from $\chi^2$ minimization.
The mean and standard deviation of these values are then calculated for the four cases of $\tau_{\rm hydro}$.
The extracted $\eta/s$ values for different switching times are given in Tab.~\ref{tab:eta_s}.
Interestingly, $\tau_{\rm hydro}$ does not significantly affect the value of $\eta/s$, such that they all agree within error bars. 
This means that this a priori free parameter of the hybrid setup does not actually influence our extracted value of the shear viscosity over entropy ratio, which is why it does not appear in Table~\ref{tab:model_parameters_list}.
Averaging the results in Tab.~\ref{tab:eta_s} yields $\langle\eta/s\rangle_{\tau_{\rm hydro}}=0.31\pm 0.05$. 
\begin{figure}[!t]
    \centering
    \includegraphics[width=\linewidth]{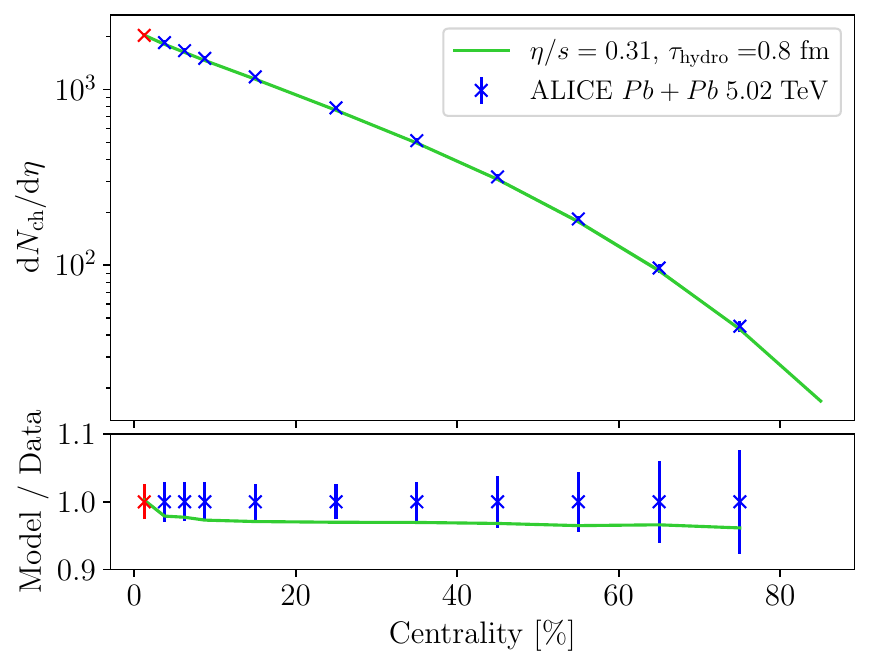}
    \caption{Charged hadron multiplicity predictions across centrality classes with the extracted $\eta/s$ value of 0.31 and $\tau_{\rm hydro}=0.8\;\mathrm{fm}$ (line), compared to ALICE data~\cite{ALICE:2015juo} (points). The red point indicates the most central collision bin used for the fit.}
    \label{fig:centrality_prediction}
\end{figure}
\begin{table}[!tb]
    \centering
    \caption{\label{tab:eta_s}Extracted $\eta/s$ values for various values of $\tau_{\rm hydro}$.}
    \begin{tabular}{c|c}
        \hline\hline
        $\tau_{\rm hydro}\;[\mathrm{fm}]$ & $\eta/s$ \\
        \hline
        $0.4$ & $0.30\pm 0.04$ \\
        $0.6$ & $0.31\pm 0.04$ \\
        $0.8$ & $0.31\pm 0.05$ \\
        $1.0$ & $0.30\pm 0.04$ \\
        \hline\hline
    \end{tabular}
\end{table}

Figure~\ref{fig:centrality_prediction} illustrates the charged hadron multiplicity at midrapidity as a function of centrality, comparing the model predictions to experimental data.
The most central class point used for the fit is highlighted in red, showing excellent agreement with the model. 
The model also provides a good description of the multiplicity in mid-central and peripheral collisions, albeit it slightly underpredicts the results in all but the most central events by a few percent. 

\subsection{Temperature extraction}
\label{subsec:T_extraction}

We can now estimate an effective temperature associated with entropy production in the pre-equilibrium stage of the heavy-ion collision evolution, to put our value of $\langle\eta/s\rangle_{\tau_{\rm hydro}}=0.31\pm 0.05$ into context with other $\eta/s$ extraction methods.

For that purpose, we use the saturation model with the MAP parameters from the DIS fit and the mean value of the $K$-factor determined in the previous sections to generate a few initial states of $\Pb+\Pb$ collisions at impact parameter $b=0$, to account for similar initial geometry and density fluctuations as used in the calibration of the multiplicity estimator formula.
Each of these initial profiles is used as input for K{\o}MP{\o}ST, which is run with the mean value of $\eta/s = 0.31$. 
For each initial profile, the K{\o}MP{\o}ST evolution is carried out up to various proper times in the range $\tau\in[0.001,1.101]\;\mathrm{fm}$ in increments of $0.001\;\mathrm{fm}$. 
At the end of this evolution,\footnote{The somewhat intricate procedure is due to the construction of K{\o}MP{\o}ST, which needs to know the total duration of the evolution to estimate which spatial region will causally affect the physics at a given point at the end of the evolution.} we estimate from the local energy density $e(\tau,\xt)$ the equivalent temperature and (equilibrium) entropy density~\cite{Kurkela:2018vqr}, using the conformal equation of state of the K{\o}MP{\o}ST system:
\begin{align}
    T(\tau,\xt) &= \left(\frac{30}{\nu_{\rm eff}\pi^2}\right)^{\!\tfrac{1}{4}}e(\tau,\xt)^{\tfrac{1}{4}} \equiv \mathcal{C}_T e(\tau,\xt)^{\tfrac{1}{4}}, \\
    s(\tau,\xt) &= \frac{4}{3}\frac{e(\tau,\xt)}{T(\tau,\xt)} = 
      \frac{4}{3_{}\mathcal{C}_T}e(\tau,\xt)^{\tfrac{3}{4}},
\end{align}
where $\mathcal{C}_T\equiv [30/(\nu_{\rm eff}\pi^2)]^{1/4}$ and $\nu_{\rm eff}=40$, consistent with the value used in the determination of $C(\eta/s)$ for the multiplicity estimator calibration in Eq.~\eqref{eq:est_calib2}.
This allows us to obtain the time dependence of the local temperature and entropy density throughout the whole pre-equilibrium stage for each chosen initial state.

From the obtained $T(\tau,\xt)$ and $s(\tau,\xt)$, we define an average effective temperature weighted by entropy production as 
\begin{equation}
\label{eq:avg_temperature}
\langle T_\mathrm{eff} \rangle|_\tau = \frac{\displaystyle\int_{\tau'_0}^\tau\d\tau'\!\int\!\d^2\xt\;T(\tau',\xt)\mathcal{S}(\tau',\xt)}{\displaystyle\int_{\tau'_0}^\tau\d\tau'\!\int\!\d^2\xt\;\mathcal{S}(\tau',\xt)},
\end{equation}
where $\mathcal{S}(\tau,\xt)\equiv \d(\tau s(\tau,\xt))/\d\tau$ denotes the entropy production rate, while the integrals over position run over the whole transverse plane.
Numerically, we compute 
\begin{align*}
    \langle T_\mathrm{eff} \rangle|_\tau \simeq \frac{\displaystyle\sum_{\xt}\sum_{\tau_i=\tau_0'}^\tau \bar{T}(\tau,\xt)\left[\tau_{i+1}s(\tau_{i+1},\xt) - \tau_is(\tau_i,\xt)\right]}{\displaystyle\sum_{\xt}[\tau s(\tau,\xt)-\tau_0' s(\tau_0',\xt)]},
\end{align*}
where $\sum_\xt$ represents a sum over the grid, and the average temperature at a transverse position between two successive time steps is defined as $\bar{T}(\tau,\xt)\equiv [T(\tau_{i+1},\xt)+T(\tau_i,\xt)]/2$.

The time integral in Eq.~\eqref{eq:avg_temperature} naturally depends on the lower bound $\tau'_0$. 
In principle, this should coincide with the initial time $\tau_0 = 0.001\;\mathrm{fm}$ of the K{\o}MP{\o}ST evolution. 
However, the early-time behavior $e\propto \tau^{-1}$ in K{\o}MP{\o}ST leads to a logarithmic divergence of $\langle T_\mathrm{eff} \rangle$ with $\tau'_0$: 
Indeed, $e\propto \tau^{-1}$ results in $\tau s\propto \tau^{1/4}$ and thus ${\cal S}\propto \tau^{-3/4}$. 
Multiplying by $T\propto \tau^{-1/4}$ yields $T{\cal S}\propto \tau^{-1}$ in the integrand of the numerator of Eq.~\eqref{eq:avg_temperature}. 
Since the denominator remains finite in the limit of small $\tau'_0$, one thus obtains the announced logarithmic divergence $\langle T_{\rm eff}\rangle\propto \ln(\tau_0^\prime)$.

Physically, K{\o}MP{\o}ST is not the right model to describe the model at such early times, at which the system is dominated by color fields. 
Thus, the lower bound of the time integrals in Eq.~\eqref{eq:avg_temperature} should be restricted to values that are at least of order $1/Q_s$. 
Since the latter is not defined precisely, we account for this uncertainty by varying $\tau_0^\prime$ between 0.08 fm and 0.18 fm to compute the mean effective temperature.

\begin{figure}[!tb]
    \centering
    \includegraphics[width=\linewidth]{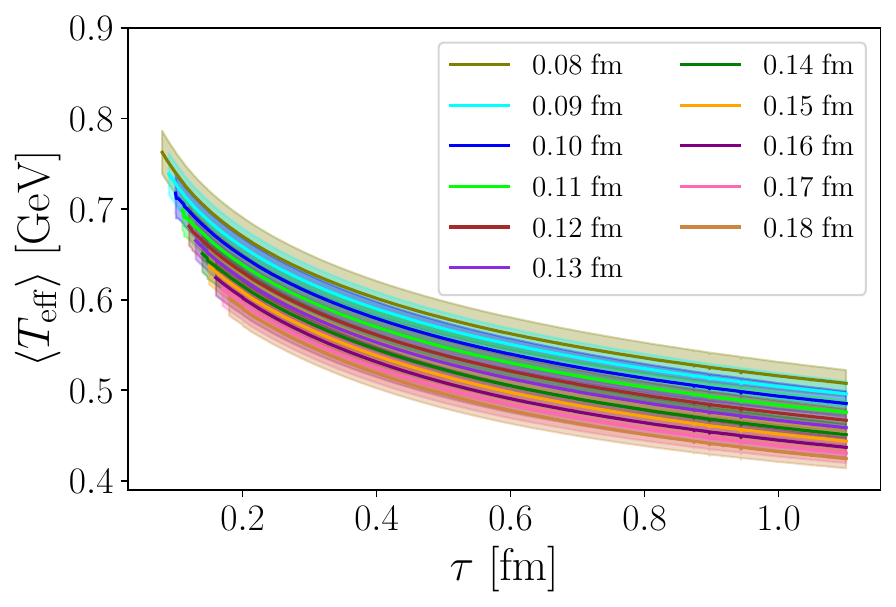}
    \caption{Effective average temperature weighted by entropy production (Eq.~\eqref{eq:avg_temperature}) as a function of proper time $\tau$ in  K{\o}MP{\o}ST averaged over five events with $b=0$. Each curve represents a different initial time $\tau_0'$ for the integration. The colored band indicates the standard deviation of the five events.}
    \label{fig:temperature}
\end{figure}

Figure~\ref{fig:temperature} shows the proper-time dependence of the effective temperature~\eqref{eq:avg_temperature}, averaged over five different initial states at $b=0$, for different starting points of the time integration.
We find that most of the entropy production occurs up to a proper time of $\tau\approx 0.6\;\mathrm{fm}$, after which the entropy-production-weighted temperature presents a slower decrease.
This observation aligns with our earlier finding that the extracted $\eta/s$ values remain consistent across the four different switching times from K{\o}MP{\o}ST to MUSIC, $\tau_{\rm hydro}\in[0.4,1.0]\;\mathrm{fm}$.
Averaging over the different starting times $\tau_0^\prime$ the values of $\langle T_\mathrm{eff} \rangle$ at the previously used $\tau_{\rm hydro}$  gives us the values displayed in Tab.~\ref{tab:T}.
\begin{table}[!tb]
    \centering
    \caption{\label{tab:T}Extracted $\langle T_{\rm eff}\rangle$ values for various values of $\tau_{\rm hydro}$. The values are an average between five different events at that $\tau_\mathrm{hydro}$, and the error bar is determined from a combined error bar in each curve and the standard deviation between the different lines in Fig.~\ref{fig:temperature}.}
    \begin{tabular}{c|c}
        \hline\hline
        $\tau_{\rm hydro}\;[\mathrm{fm}]$ & $\langle T_{\rm eff}\rangle\;[\mathrm{MeV}]$ \\
        \hline
        $0.4$ & $556\pm 32$ \\
        $0.6$ & $505\pm 37$ \\
        $0.8$ & $489\pm 31$ \\
        $1.0$ & $469\pm 30$ \\
        \hline\hline
    \end{tabular}
\end{table}
\begin{figure*}[!tb]
    \centering
    \includegraphics[width=\linewidth]{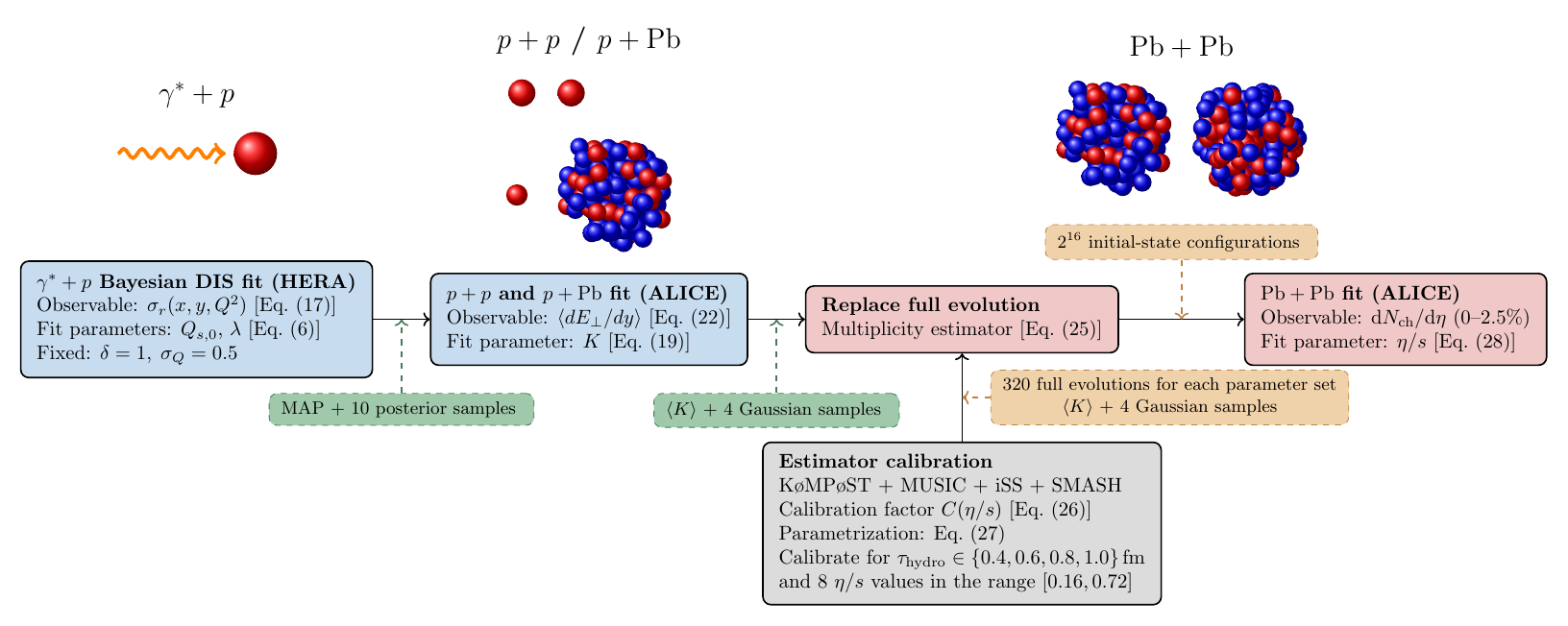}
    \caption{Illustration of the fitting routine from Sec.~\ref{sec:global_fit}. 
    Cold nuclear matter fits are in blue, the hot nuclear matter fit is in red, and the full evolution is replaced by a multiplicity estimator calibration (gray). 
    Propagated parameter sets for uncertainty estimation are shown in green.}
    \label{fig:fitting_scheme}
\end{figure*}
Accordingly, we obtain a range for the typical temperature scale at which entropy is produced by dissipative effects in the pre-equilibrium stage. 
Since the conversion from initial energy to final entropy, as measured by the charged multiplicity, mostly happens in this stage, this is also the typical temperature scale to which the extracted shear viscosity over entropy ratio $\eta/s$ corresponds.

\section{Conclusions}
\label{sec:conclusions}

We performed a proof-of-principle study, to simultaneously constrain the properties of hot and cold nuclear matter within a saturation physics-based model for computing DIS cross-section, transverse energy in $p+p$, $p+A$, and charged particle multiplicities $A+A$ collisions.\footnote{In an early study~\cite{Armesto:2004ud}, data from DIS on protons and heavier nuclei were used to make predictions, based on geometrical scaling, for the multiplicity in $A+A$ collisions across a range of center-of-mass energies. Yet the model does not involve any medium-specific parameter, as $\eta/s$ in the present study, but instead relies on an overall normalization constant for the multiplicity, which is deduced from the data.}
The fitting procedure is schematically displayed in Fig.~\ref{fig:fitting_scheme}.
We first constrained certain model parameters ($Q_{s,0}$, $\lambda$) using HERA data from DIS, then calibrated the $K$-factor for $p+p/p+A$ systems.
We demonstrated how to convert the charged-particle multiplicity probability distribution into one for the initial-state transverse energy $\d E_\perp/\d y$ and used its mean value to determine the $K$-factor within the saturation model.

In the final step, we calibrated an estimator formula to determine the final-state charged hadron multiplicity from the initial-state energy density profile, assuming that most of the entropy density is produced during the pre-equilibrium evolution.
This approach enabled the efficient generation of initial-state profiles with direct multiplicity estimates, allowing us to extract $\eta/s$.
Our final result, $\langle\eta/s\rangle_{\tau_{\rm hydro}}=0.31\pm 0.05$, was found to be insensitive to the switching time between pre-equilibrium and hydrodynamics.
The method presented here offers a computationally efficient way to determine the shear viscosity of the non-equilibrium early-time QGP based solely on the observed multiplicity.
Additionally, we extracted an effective temperature $\langle T_{\rm eff}\rangle$ for the pre-equilibrium stage, facilitating comparison with other methods.

\begin{figure}[!tb]
    \centering
    \includegraphics[width=\linewidth]{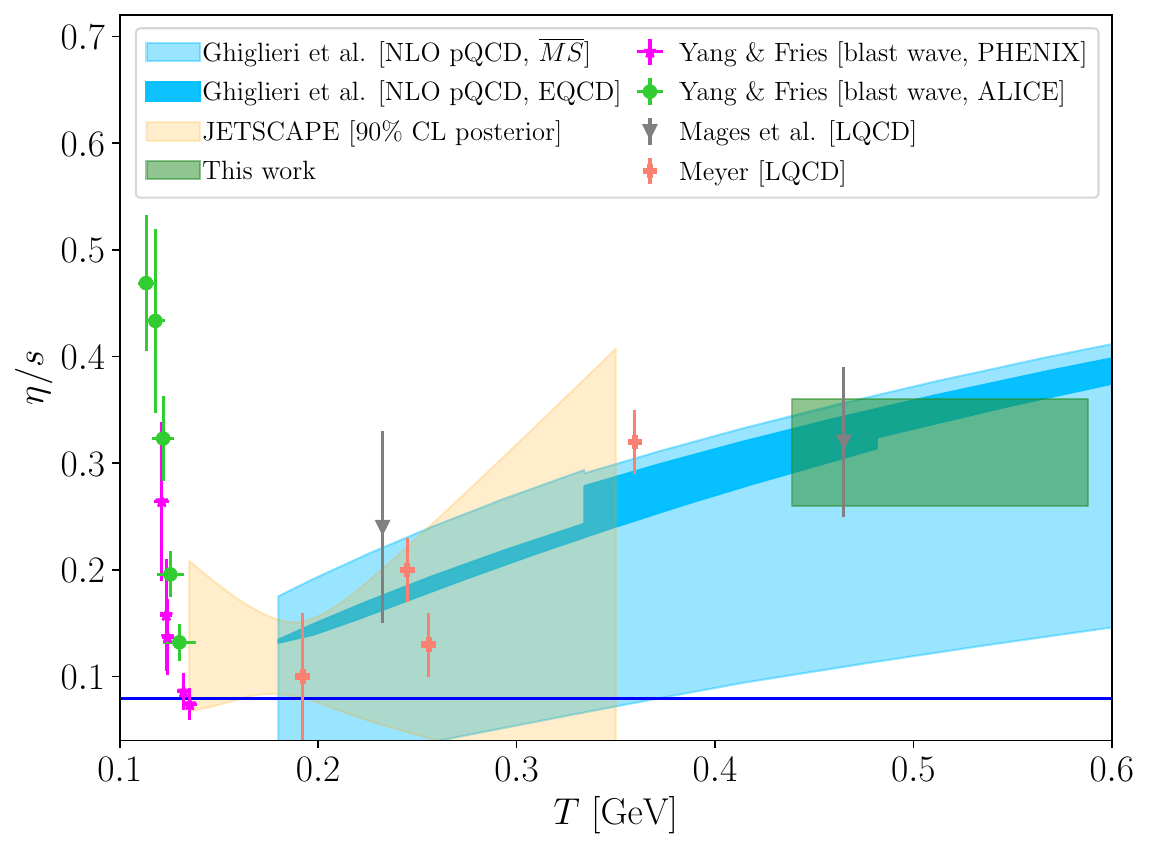}
    \caption{Comparison of the $\eta/s$ value extracted in this work (green band) with values from the literature obtained using different techniques~\cite{Meyer:2007ic,Meyer:2009jp,Mages:2015rea,Ghiglieri:2018dib,Yang:2022yxa,JETSCAPE:2020shq}.}
    \label{fig:eta_s_vs_T}
\end{figure}

Figure~\ref{fig:eta_s_vs_T} presents the final result for $\eta/s$ in the pre-equilibrium phase of a heavy-ion collision, extracted at the temperature obtained from the K{\o}MP{\o}ST evolution.
We compare our result (green area) with other values for $\eta/s(T)$ from several other approaches~\cite{Meyer:2007ic,Meyer:2009jp,Mages:2015rea,Ghiglieri:2018dib,Yang:2022yxa,JETSCAPE:2020shq}.
The green area extracted from this work corresponds to a $1\sigma$ region for the minimum and maximum of the extracted $\eta/s$ and $\langle T_{\rm eff}\rangle$ values.
Our value of $\eta/s(\langle T_{\rm eff}\rangle)$ is in agreement with the values obtained from lattice QCD (LQCD) and perturbative QCD (pQCD) at NLO.
The figure also suggests a good agreement with what extending the posterior band of the Bayesian analysis from~\cite{JETSCAPE:2020shq} to larger temperatures would yield, although it should be mentioned that the experimental dataset used in that analysis does not constrain this region when the posterior distribution is compared to the prior (not shown).

Let us emphasize that the comparison of Fig.~\ref{fig:eta_s_vs_T} is not meant to claim direct agreement, but rather to show that our result falls within a reasonable range relative to other approaches.
The viscosity in our work is extracted from the pre-equilibrium phase, where only an effective temperature can be defined.
At this stage, the system has not yet reached thermodynamic equilibrium, meaning that our value represents only an upper bound for $\eta/s$ at such temperatures, whereas the pQCD and LQCD calculations are performed in equilibrium.

This exploratory multistage fit shows how the pre-equilibrium $\eta/s$ can be extracted with the help of experimental data by fixing relevant model parameters step by step.
We have shown that this procedure can put tight constraints on the $\eta/s$ at high temperatures, where Bayesian inference studies typically do not have much constraining power.
However, we also note that the final extracted value is highly sensitive to the value of $\langle\mathrm{d}E_\perp/\mathrm{d}y\rangle$ in the $p+p$ and $p+\Pb$ fit, which makes precise experimental measurements of this quantity important to fix the $K$ factor in initial-state models.

Comparing our results with the Bayesian extraction of $\eta/s$ from Ref.~\cite{Heffernan:2023utr} (not shown in Fig.~\ref{fig:eta_s_vs_T}), which also employs a CGC-based initial-state model, we note important differences in the treatment of the pre-equilibrium phase. 
In Ref.~\cite{Heffernan:2023utr}, the IP-Glasma initial conditions are directly matched to hydrodynamics after the Glasma evolution, whereas in the present work we use the GBW model as input for a kinetic-theory pre-equilibrium evolution in K{\o}MP{\o}ST. 
This difference in early-time dynamics likely explains the qualitatively different temperature dependence of $\eta/s$: in our study, $\eta/s$ is larger at the effective temperature of the early-time evolution, in agreement with pQCD and lattice QCD, whereas the Bayesian posterior in Ref.~\cite{Heffernan:2023utr} shows a decreasing trend with increasing temperature. 
These observations highlight the sensitivity of the extracted $\eta/s$ to the modeling of the pre-equilibrium phase.

While our results clearly demonstrate that rather stringent constraints on QGP transport properties can be derived from simple multiplicity measurements in $A+A$ collisions, as long as the energy density in the initial state is sufficiently constrained, the exploratory study in this work only represents a first step in this regard, and there are several ways in which the analysis can be improved. 
For instance, future work could extend this study by incorporating the full transverse energy probability distribution into the fitting procedure.
This would require a combined fit of DIS data and $p+p/p+A$ systems, as the parameter $\sigma_Q$ appears in both cases.
Such an approach would be computationally more demanding, and it may be beneficial to expand the small-scale Bayesian calibration performed for the DIS data to additional steps in the fitting process demonstrated in this work.
In this case, it would also be important to include sub-nucleonic degrees of freedom in the fit to improve the description of the transverse energy distribution.
Evidently it would be desirable to extend the procedure outlined in this work to a fully fledged Bayesian analysis of heavy-ion collisions (see e.g.\ Refs.~\cite{Novak:2013bqa,Bernhard:2016tnd,Nijs:2020ors,JETSCAPE:2020shq}), which include temperature dependent shear and bulk viscosities, but usually treat the initial state energy density as a free model parameter, thus effectively ignoring the important constraints from multiplicity measurements.

We stress that in the present work the full CGC+K{\o}MP{\o}ST+MUSIC+SMASH evolution is only used to calibrate the multiplicity estimator~\eqref{eq:mult_estimator}-\eqref{eq:C_estimator_prefac}. 
Once this calibration is performed, the initial-state energy density profiles are sufficient to determine the charged-particle multiplicity efficiently, without running the full dynamical evolution. 
Consequently, flow observables such as $v_2$ cannot be computed in this setup, as they require event-by-event evolution from the initial to the final hadronic state. 
While the shear viscosity values used in this study are physically reasonable for the early-time QGP, they may differ from those typically needed to reproduce flow observables in hydrodynamic simulations, which are sensitive to later-time evolution at intermediate temperatures. 

Looking forward, the methodology presented here could be extended to a full Bayesian analysis of heavy-ion collisions, including flow observables and a temperature-dependent $\eta/s(T)$ and $\zeta/s(T)$, while retaining constraints from DIS and small-system data. 
In this way, the framework demonstrated here provides a path toward a comprehensive, state-of-the-art extraction of QGP transport properties across all stages of the collision, while fully accounting for correlations between initial- and final-state observables.

\textbf{Data Availability Statement}. Ancillary files with data of the figures are appended to the arXiv submission, \href{https://doi.org/10.48550/arXiv.2504.02726}{arXiv:2504.02726}.

\textbf{Code Availability Statement.} Part of the codes used in the current study is referred to in Refs.~\cite{hendrik_roch_2024_12807892,hendrik_roch_2024_12694840,hendrik_roch_2024_14516063}; the rest is available from the authors on reasonable request.

\begin{acknowledgments}
We thank Rainer J. Fries, Jacopo Ghiglieri, Guy D. Moore, Jean-Fran\c{c}ois Paquet, and Zhidong Yang for providing the data in Fig.~\ref{fig:eta_s_vs_T}, and Derek Teaney for discussions.
This work is supported by the programme Netz\-werke 2021, an initiative of the Ministry of Culture and Science of the State of Northrhine Westphalia (MKW NRW)
under the funding code NW21-024-A.
The sole responsibility for the content of this publication lies with the authors.
N.~B., R.~K. and S.~S. acknowledge support by the Deutsche Forschungsgemeinschaft (DFG, German Research Foundation) through the CRC-TR 211 'Strong-interaction matter under extreme conditions' - project number 315477589 - TRR 211.
H.~R. was supported by the National Science Foundation (NSF) within the framework of the JETSCAPE collaboration (OAC-2004571) and by the DOE (DE-SC0024232).
X.~D. acknowledges support by the European Research Council project ERC-2018-ADG-835105 YoctoLHC, Spanish Research State Agency under project PID2020-119632GB-I00, Xunta de Galicia (Centro singular de investigacion de Galicia accreditation 2019-2022), European Union ERDF.
Numerical simulations presented in this work were performed at the Paderborn Center for Parallel Computing (PC$^2$) and we gratefully acknowledge their support.
\end{acknowledgments}

\appendix
\section{Self consistent solution of nuclear saturation scale}
\label{app:self_consistent_Qs}

\begin{figure}[!t]
    \centering
    \includegraphics[width=0.40\textwidth]{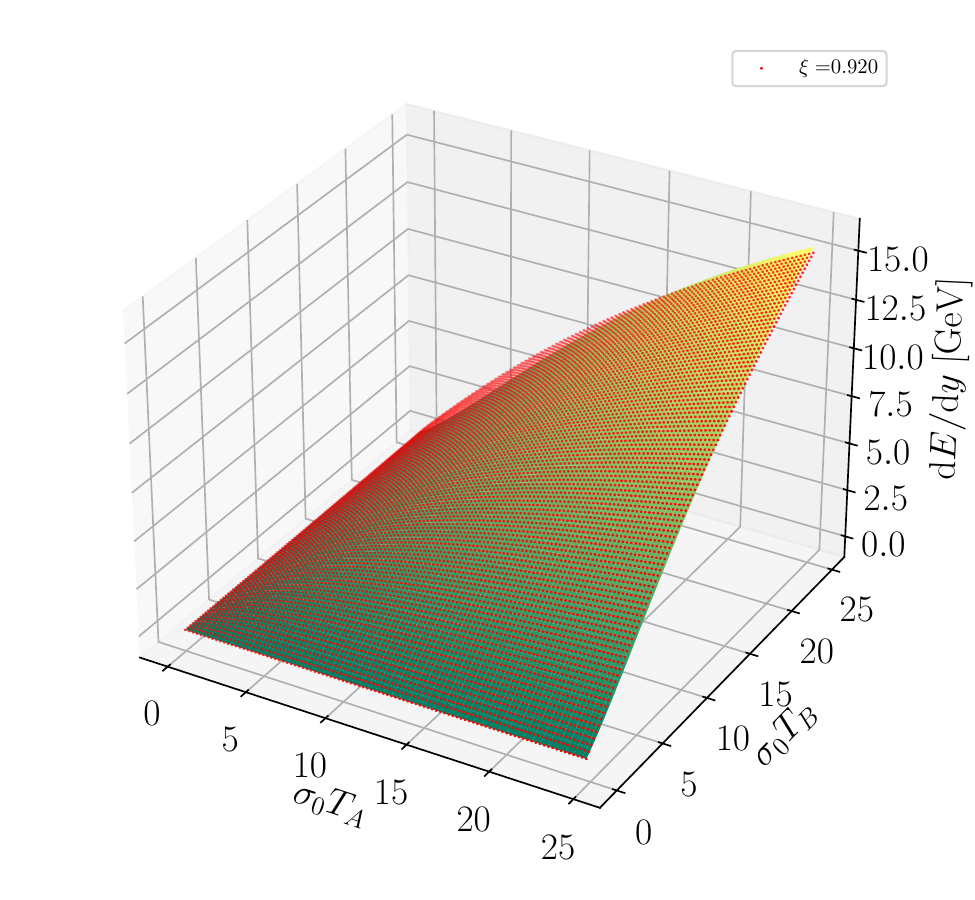}
    \caption{$\d E/\d y$ numerically integrated using Eqs.~\eqref{eq:NgSpectra}-\eqref{eq:energyperrapidity} (surface) and the analytical treatment making use of approximation~\eqref{eq:QsXSelfConsistent} (red dots) for different values of $\sigma_0T_{A/B}$. The approximation is optimized via the factor $\xi$ in Eq.~\eqref{eq:x_approx}.}
    \label{fig:analytical_vs_numerical}
\end{figure}
The full numerical solution of Eqs.~\eqref{eq:NgSpectra}-\eqref{eq:energyperrapidity} without approximating the $x$-dependence of the UGDs and the analytical treatment of the integrals using Eq.~\eqref{eq:QsXSelfConsistent} can give different results for the initial energy per unit rapidity $\d E/\d y$.
This is why we allow for an order 1 factor $\xi$ in Eq.~\eqref{eq:QsXSelfConsistent} to correct the uncertainty of the approximation:
\begin{align}
x_{A/B} = \xi\cdot\frac{Q_{s,A/B}(x_{A/B},\xt)\,{\rm e}^{\pm Y}}{\snn}.
\label{eq:x_approx}
\end{align}
To find the optimal value of $\xi$ we compute the full numerical integral for the initial energy per unit rapidity for different values of $\sigma_0 T_A$ and $\sigma_0 T_B$ and minimize the difference between the full solution and the approximation.
The comparison of the two methods for the optimized $\xi$ value is shown in Fig.~\ref{fig:analytical_vs_numerical}.
Adopting the parameters $Q_{s,0}$, $\lambda$, $\delta$ and $\sigma_Q$ from the DIS fit (see Sec.~\ref{subsec:DIS_fit}), we find $\xi=0.920$ as the optimal choice for obtaining the initial state of hadronic collisions at $\sqrt{s_{\rm NN}}=5.02\;\mathrm{TeV}$, with a very weak dependence on the energy of the collisions at LHC energies: for $\sqrt{s_{\rm NN}}=2.76\;\mathrm{TeV}$ the optimal $\xi$ value is 0.921.

Note that, similar to the constants entering the multiplicity estimator~\eqref{eq:mult_estimator}--\eqref{eq:C_estimator_prefac}, $\xi$ is not to be viewed as a parameter of the model. 
Instead, it is a parameter in an approximate formula, Eq.~\eqref{eq:x_approx}, that replaces the more time- and resource-demanding evaluation of integrals, in order to accelerate the computation of one of the model ingredients.

\section{Transverse energy distribution}
\label{app:ET_dist}

In Ref.~\cite{ALICE:2022imr}, the ALICE collaboration introduced a method to estimate the average transverse-energy rapidity density, $\langle\d E_\perp/\d y\rangle$, using the relation:
\begin{align}
    \left\langle\frac{\d E_\perp}{\d y}\right\rangle &\simeq \frac{\langle m\rangle}{f_{\rm total}} \sqrt{1+a^2}\,\frac{\d N_{\rm ch}}{\d y}\nonumber\\
    &=\frac{\langle m\rangle}{f_{\rm total}} \sqrt{1+a^2}\sqrt{1+\frac{1}{a^2}\frac{1}{\cosh^2\eta}}\frac{\d N_{\rm ch}}{\d\eta}
    \label{eq:ET_ALICE}
\end{align}
with $a\equiv \langle p_{\rm T}\rangle/\langle m\rangle$.
Here, $\langle m\rangle=(0.215\pm 0.001)\;\mathrm{GeV}$ is the average hadron mass based on identified-hadrons data~\cite{ALICE:2013mez} and $f_{\rm total}=0.55\pm 0.01$ is the fraction of charged hadrons~\cite{ALICE:2016igk}, as provided by ALICE~\cite{ALICE:2022imr}. The formula accounts for the contribution of neutral particles, which are not measured directly, to the transverse energy.

Since the saturation model predicts the total available transverse energy, the charged multiplicity used in Eq.~\eqref{eq:ET_ALICE} should include the full transverse-momentum space. 
Thus, we extrapolate the charged-hadron multiplicity distribution measured by ALICE~\cite{ALICE:2022xip} in the kinematic range $\abs{\eta}<0.8$ and $0.15\;\mathrm{GeV}<p_{\rm T}< 10\;\mathrm{GeV}$ to the full momentum range. 
Three fit functions (L\'evy-Tsallis~\cite{Tsallis:1987eu,STAR:2006nmo}, modified-Hagedorn\,\cite{Hagedorn:1983wk}, two-component model\,\cite{Bylinkin:2014qea}) are employed to extrapolate the published spectra to $p_{\rm T}=0$ to obtain $\d N_{\rm ch}/\d\eta$ and $\langle p_{\rm T}\rangle$.
To take into account uncertainties in the parameters, we perform 1000 Gaussian samples around the mean values for $\langle m\rangle$ and $f_{\rm total}$ for each fit method.
At the end, the values of charged multiplicity and mean transverse momentum from the three fit methods are averaged. 
These values are used to compute the average transverse-energy density~\eqref{eq:ET_ALICE}, with which we calibrate the $K$-factor of the saturation model in Sec.~\ref{subsec:pp_pA_fit}.
Since the values are highly correlated, we did not include Gaussian samples for the transverse momentum and charged hadron multiplicity.

To determine the probability distribution for the transverse energy density $p(\mathrm{d}E_\perp/\mathrm{d}y)$ we use the experimental ALICE data on the $N_{\rm ch}$ dependence of $\langle p_{\rm T}\rangle$ and the probability density of $N_{\rm ch}$~\cite{ALICE:2022imr}.
We perform a cubic spline interpolation of $\langle p_{\rm T}\rangle$ and $p(N_{\rm ch})$, using only every third experimental data point to avoid wiggles in the final distribution, especially in the tail of the distribution.
In the next step we use Eq.~\eqref{eq:ET_ALICE} to determine $\mathrm{d}E_\perp/\mathrm{d}y$ for each bin in $N_{\rm ch}$.
This transverse-energy density is then interpolated with a cubic spline as a function of $N_{\rm ch}$, from which we compute the derivative $\d E_\perp/\d N_{\rm ch}$.
Eventually, we divide the interpolated $p(N_{\rm ch})$ by the derivative to obtain $p(\mathrm{d}E_\perp/\mathrm{d}y)$.
To estimate the uncertainty on this distribution, we perform 1000 Gaussian samples about the mean values of $\langle m\rangle$ and $f_{\rm tot}$ in Eq.~\eqref{eq:ET_ALICE}.




\bibliography{ref.bib}

\end{document}